\documentclass[aip,jcp]{revtex4-1}

\usepackage[latin1]{inputenc}
\usepackage[T1]{fontenc}
\usepackage{graphicx}
\usepackage{amsmath}
\usepackage{amssymb}
\usepackage{color}

\def\ba{\begin{align}}
\def\ea{\end{align}}
\newcommand{\brv}{(\mathbf{r})}
\newcommand{\rv}{\mathbf{r}}
\newcommand{\fv}{\mathbf{f}}
\newcommand{\td}{\text{d}}
\newcommand{\xc}{exchange-correlation~}

\begin{document}

	\title{Energy Density Functionals From the Strong-Coupling Limit Applied to the Anions of the He Isoelectronic Series}
	
	\author{Andr\'e Mirtschink}
	\affiliation{Department of Theoretical Chemistry and Amsterdam Center for Multiscale Modeling, FEW, Vrije Universiteit, De Boelelaan 1083, 1081HV Amsterdam, The Netherlands}
	\author{C. J. Umrigar}
	\affiliation{Laboratory of Atomic and Solid State Physics, Cornell University, Ithaca, New York 14853, USA}
	\author{John D. Morgan III}
	\affiliation{Department of Physics and Astronomy, University of Delaware, Newark, DE 19716, USA}
	\author{Paola Gori-Giorgi}
	\affiliation{Department of Theoretical Chemistry and Amsterdam Center for Multiscale Modeling, FEW, Vrije Universiteit, De Boelelaan 1083, 1081HV Amsterdam, The Netherlands}
	\date{\today}
	
\begin{abstract}

Anions and radicals are important for many applications including environmental chemistry, semiconductors, and charge transfer, but are poorly described by the available approximate energy density functionals. Here we test an approximate exchange-correlation functional based on the exact strong-coupling limit of the Hohenberg-Kohn functional on the prototypical case of the He isoelectronic series with varying nuclear charge $Z<2$, which includes weakly bound negative ions and a quantum phase transition at a critical value of $Z$, representing a big challenge for density functional theory. We use accurate wavefunction calculations to validate our results, comparing energies and Kohn-Sham potentials, thus also providing useful reference data close to and at the quantum phase transition. We show that our functional is able to bind H$^-$ and to capture in general the physics of loosely bound anions, with a tendency to strongly overbind that can be proven mathematically. We also include corrections based on the uniform electron gas which improve the results.

\end{abstract}

\keywords{Density Functional Theory, Energy Density Functional, Exchange-Correlation, Strong Interaction Limit, Strong Coupling Limit, Strictly Correlated Electrons, Helium Isoelectronic Series}

	\maketitle

\section{Introduction}
Density functional theory (DFT),\cite{HohKoh-PR-64} in its Kohn-Sham (KS) formulation,\cite{KohSha-PR-65} has been a real breakthrough for electronic structure calculations. The key idea of KS DFT is an exact mapping\cite{KohSha-PR-65} between the physical, interacting, many-electron system and a model system of non-interacting fermions with the same density, allowing for a realistic treatment of the electronic kinetic energy. All the complicated many-body effects are embedded in the so-called exchange-correlation (xc) energy functional.
Although, in principle, the exact xc functional is unique (or ``universal''), in practice a large number of approximations has been developed in the last thirty years, often targeting different systems, different properties, and different phenomena. Common practice for DFT users is nowadays to consult the (rather extensive) benchmark literature to choose the approximate xc functional most suitable for the problem at hand.
This reflects the intrinsic difficulty of building a general approximation able to recognize and capture, for each class of systems or process, the many-body effects relevant for its description.

Even in this ``specialized-functional'' world, there are still important cases in which state-of-the-art KS DFT encounters severe problems, which is why the quest for better xc functionals continues to be a very active research field (for a recent review, see, e.g., Ref.~\onlinecite{CohMorYan-CR-12}). The most notable example is the treatment of near-degeneracy and strong-correlation effects, which involve rearrangement of electrons within partially filled levels. These effects appear in bond dissociation but also at equilibrium geometries, particularly for systems with $d$ and $f$ unsaturated shells, such as transition metals and actinides. Mott insulators and low-density nanodevices are other examples of strongly-correlated  systems whose physics is not captured by the standard approximations. A key problem when dealing with strong (or ``static'') correlation is that, similarly to unrestricted Hartree-Fock, approximate KS DFT tries to mimic the physics of strong correlation and near degeneracy with spin and spatial symmetry breaking, which in complex systems may occur erratically and can be very sensitive to the choice of functional.\cite{CraTru-PCCP-09} This  easily leads to a wrong characterization of several properties and to discontinuous potential energy surfaces.\cite{CraTru-PCCP-09} Being able to capture strong electronic correlation within KS DFT without resorting to symmetry breaking is arguably one of the most important problems of electronic structure theory.\cite{Bec-JCP-13a,Bec-JCP-13b,Bec-JCP-13c,CohMorYan-CR-12,CraTru-PCCP-09}

The mainstream strategies to construct approximate functionals consist of making an ansatz for the dependence of the xc functional on the relevant ``ingredients'' such as the local density, the local density gradients, the KS kinetic orbital energy, the KS orbitals, etc.\cite{PerRuzTaoStaScuCso-JCP-05} The ansatz can be constructed in order to fulfill as many exact constraints as possible given the ingredients used.\cite{PerRuzTaoStaScuCso-JCP-05} Some authors also introduce a (sometimes very large) number of parameters to be fitted to a specific data set (for recent reviews, see, {\it e.g.}, Refs.~\onlinecite{CraTru-PCCP-09,CohMorYan-CR-12,PevTru-PTRSA-14}).

In recent years, an exact piece of information on the exact exchange-correlation functional, namely the limit of infinite correlation,\cite{SeiGorSav-PRA-07,GorVigSei-JCTC-09,GorSeiVig-PRL-09} has become available. The ``strictly-correlated-electrons'' (SCE) functional, that utilizes this information, has a highly non-local dependence on the density, but its functional derivative (yielding the KS potential) can be easily constructed via a rigorous and physically transparent shortcut.\cite{MalGor-PRL-12,MalMirCreReiGor-PRB-13} The SCE functional becomes exact in the limit in which the electron-electron interaction dominates over the electronic kinetic energy, and it has been successfully applied to model low-density quantum wires\cite{MalGor-PRL-12,MalMirCreReiGor-PRB-13} and quantum dots.\cite{MenMalGor-PRB-14} In those systems, the SCE functional has been shown capable of capturing the physics of charge localization without introducing magnetic order or any other symmetry breaking. In other words, the SCE functional achieved what was often regarded as practically impossible: making non-interacting electrons behave as strongly-correlated ones, showing that restricted KS DFT with the appropriate functionals can yield results beyond mean-field theory.

Using the SCE functional to address chemical problems also seems very attractive. It provides a new, well defined, starting point to build approximate functionals, deeply different from mainstream approaches. The new ingredient here is the non-locality encoded in the SCE functional and potential, which can capture the physics that is missed by standard approximations.
Chemistry, however, is more challenging for the SCE functional than low-density nanostructures, because the kinetic energy and the electron-electron repulsion often have similar importance. For example, in a stretched bond only the bonding electrons are strongly-correlated, while the others are not. Indeed, in a recent paper,\cite{MalMirGieWagGor-PCCP-14} it has been shown that KS SCE dissociates properly a single chemical bond without introducing symmetry breaking, but it overcorrelates in all other aspects.
This evidently requires corrections to the SCE functional, which can be built either by including higher-order terms in the expansion at infinite coupling strength\cite{GorVigSei-JCTC-09} or by considering rigorous local and semilocal approximations.\cite{MalMirCreReiGor-PRB-13}

Both low-density nanostructures and stretched bonds involve charge localization due to strong spatial correlations, which is, by definition, the case in which SCE tends asymptotically to the exact xc functional. To gain insight into the performance of the SCE functional for other classes of chemical systems, we consider here a conceptually simple problem in which electronic correlation plays a crucial role. Despite its simplicity the problem nonetheless is very challenging for both DFT and other approaches.  This is the anions of the He isoelectronic series, described by the Hamiltonian (in Hartree atomic units used throughout the paper)
	\begin{align}
		\hat{H}=-\frac{1}{2}\nabla^2_1-\frac{1}{2}\nabla^2_2-\frac{Z}{r_1}-\frac{Z}{r_2}+\frac{1}{r_{12}},
		\label{eq:HamHes}
	\end{align}
with $Z<2$. Accurate wavefunction calculations\cite{BakFreNydMor-PRA-90} have shown  that when the nuclear charge $Z$ is lowered and crosses a critical value, $Z_{\rm crit}\approx 0.91103$, a quantum phase transition occurs from a bound to an unbound two-electron system. Thus, with this simple hamiltonian  we can explore a whole class of very loosely bound anions, including the quantum phase transition at $Z_{\rm crit}$.

As is well known, anions are problematic for state-of-the-art KS DFT. Standard approximations often yield a positive eigenvalue for the highest occupied molecular orbital (HOMO), corresponding to a quasi-bound state (or resonance) instead of a properly bound system.  Often, in practice, anions are tackled by finite basis sets within approximate DFT. Estimates of electron affinities are then obtained by the energy difference $E_{N+1}-E_N$ ($N$ being the number of electrons), ignoring the fact that the HOMO has a positive eigenvalue.\cite{MorSch-JCP-96} In the complete basis set limit (here by inclusion of plane waves), a positive orbital eigenvalue would lead to an unbound electron extending over the entire space and, consequently, within the finite basis set, orbitals with positive orbital energies should not be occupied. In practice, however, convergence of these calculations can only be achieved if the orbital with positive eigenvalue is occupied, corresponding to an electron artificially bound by the finite basis, a procedure which has been criticized.\cite{RosTri-JCP-97}

It is worth mentioning that the failure of standard DFT approximations to bind anions properly is often attributed to the self-interaction error (SIE). However, despite being self-interaction free, the Hartree-Fock (HF) method fails for H$^-$, yielding a negative binding energy for the second electron in contradiction to experiment.\cite{HotLin-JCPRD-75} Thus in this case, it is correlation that stabilizes the system, so SIE is not the only problem.

In this work we test the KS SCE functional for the hamiltonian of Eq.~(\ref{eq:HamHes}), focusing on the anions close to the quantum phase transition. We compare our results (including energies, densities and KS potentials) with those from a very accurate wavefunction treatment and from standard approximate xc functionals. We also consider local corrections to KS SCE, and we analyze some exact properties of the density and of the KS potential at $Z_{\rm crit}$.

\section{Theoretical Methods}\label{sec:methods}

\subsection{Variational Calculations from Accurate Wavefunctions}
\label{sec:wavef}
Very accurate energies of the He isoelectronic series with nuclear charge $Z$ between one and ten,\cite{FreHuxMor-PRA-84}
and for weakly bound anions close to and at the quantum phase transition,\cite{BakFreNydMor-PRA-90} have been obtained
using basis functions that depend explicitly on the interelectronic coordinates.
For the latter the wavefunction was a linear combination of 476 basis functions
consisting of 244 modified~\cite{FreHuxMor-PRA-84} Frankowski-Pekeris\cite{FraPek-PR-66} basis functions
$\phi^{{\rm FP}}_{n,l,m,j}(2Zks, 2Zkt, 2Zku)$, where
\begin{equation}
\phi^{{\rm FP}}_{n,l,m,j}(s, t, u) = s^n t^l u^m (\ln s)^j e^{-s/2},
\end{equation}
and 232 Frankowski\cite{Fra-PR-67} basis functions
$\phi^{{\rm F}}_{n,l,m,j}(2Zks, 2Zkt, 2Zku)$, where
\begin{equation}
\phi^{{\rm F}}_{n,l,m,j}(s, t, u) = s^n t^l u^m (\ln s)^j
\left(e^{ct} \pm e^{-ct}\right) e^{-s/2}.
\end{equation}
Here $k$ and $c$ are flexible scaling parameters and $s$, $t$, and $u$
are the Hylleraas coordinates
\begin{equation}
 s = r_1 + r_2, \quad t = r_2 - r_1, \quad u = r_{12}.
\end{equation}
The $\pm$ sign depends on whether $l$ is even or odd to assure the proper
symmetry of the basis functions under exchange of the two electrons
($t \, \rightarrow \, -t$). The powers $n$, $l$, $m$, and $j$ are chosen to duplicate the first several leading terms in the behavior of the exact wavefunction of a helium-like ion near the 3-particle coalescence, which is given by the Fock expansion.\cite{Foc-IANSSSR-54,Foc-proc-58,Mor-TCA-86}  
This composite basis was used in Ref.~\onlinecite{BakFreNydMor-PRA-90} to obtain compact and highly
accurate representations of the wavefunction 
of the sole bound state of the helium isoelectronic
sequence for values of $Z$ between $Z_{\rm crit} \simeq 0.9110289$ and 1.
Table~\ref{tab:valsck} shows the approximately optimal values of $k$ and $c$
used for several values of $Z$ in the present paper.

\begin{table}
\caption{Approximately optimal values of $k$ and $c$ used for several values of $Z$ in the accurate wavefunctions. The first row is from Ref.~\onlinecite{BakFreNydMor-PRA-90}.}
\begin{tabular}{lll}
\hline
   $Z$    &  $k$    & $c$  \\ \hline
0.9110289 & 0.60672 & 0.448 \\
0.92      & 0.67    & 0.41 \\
0.93      & 0.68    & 0.40 \\
0.94      & 0.69    & 0.39 \\
0.95      & 0.70    & 0.38 \\
\hline
\end{tabular}
\label{tab:valsck}
\end{table}

\subsection{Restricted KS DFT with local, semilocal, and hybrid functionals}
We quickly review some basic aspects of Kohn-Sham density functional theory, as this helps in clarifying the concepts behind the less familiar SCE functional, introduced in the next subsection.

For any $N$-electron system in the external potential $\hat{V}_{\rm ext}=\sum_{i=1}^Nv_{\rm ext}(\rv)$,  Hohenberg and Kohn (HK) have proven\cite{HohKoh-PR-64} the existence of a ``universal'' density functional $F[\rho]$, which in Levy's constrained minimization formalism\cite{Lev-PNAS-79} is
   \begin{align}
		F[\rho]=\min_{\Psi\rightarrow\rho}\left\langle\Psi | \hat{T} + \hat{V}_{ee}| \Psi \right\rangle,
		\label{eq:UF}
	\end{align}
where ``$\Psi\to\rho$'' means that the minimization is carried over all fermionic wavefunctions yielding the same one-electron density $\rho(\rv)$, so that the ground-state energy can be obtained by minimizing the energy functional
\begin{equation}
    E_0= \min_{\rho}\left\{F[\rho]+\int v_{\rm ext}(\rv)\rho(\rv)d\rv\right\}.
\end{equation}
Since it is extremely difficult to construct approximations for $F[\rho]$ that encode the fermionic nature of the electrons, Kohn and Sham have introduced another functional, the non-interacting KS kinetic energy functional,
\begin{align}
		T_s[\rho]=\min_{\Psi\rightarrow\rho}\left\langle\Psi | \hat{T} | \Psi \right\rangle,
		\label{eq:Ts}
	\end{align}
which defines  a non-interacting system of fermions with the same density of the physical, interacting, one.
The HK functional is then partitioned as
\begin{align}
		F[\rho]=T_s[\rho] + U[\rho] + E_{xc}[\rho],
		\label{eq:FKS}
	\end{align}
where  $U[\rho]$ is the classical Hartree energy, and the exchange-correlation energy $E_{xc}[\rho]$ is defined as the correction needed to make Eq.~\eqref{eq:FKS} exact. With the non-interacting KS system, the full minimization for the many-electron energy becomes equivalent to the solution of the KS one-particle equations
	\begin{align}
		\left[-\frac{1}{2}\nabla^2+v_{\rm KS}\brv\right]\psi_i\brv=\epsilon_i\psi_i\brv,
	\end{align}
with the KS potential $v_{\rm KS}$ given by the functional derivatives
	\begin{align}
		v_{\rm KS}\brv&=v_{ext}\brv+\frac{\delta U[\rho]}{\delta\rho\brv}+\frac{\delta E_{xc}[\rho]}{\delta\rho\brv}\notag\\
					&\equiv v_{ext}\brv+u_{H}\brv+v_{xc}\brv,
	\end{align}
where $u_{H}$ is the Hartree potential and $v_{xc}$ is the \xc potential. The density is obtained as $\rho(\rv)=\sum_i|\psi_i(\rv)|^2$, with the sum running over the occupied orbitals, and the KS equations are solved self-consistently.

The simplest approximation for $E_{xc}[\rho]$ is the local density approximation (LDA), defining the \xc energy as a functional of the local density alone. The next levels of refinement are the generalized-gradient approximations (GGA), obtained by including the gradient of the local density $\nabla\rho$,  and the meta-GGA functionals which use also the local Laplacian of the density $\nabla^2\rho$ and/or the local kinetic energy density $\tau(\rv)=\sum_i|\nabla \psi_i(\rv)|^2$. For the special case of the two-electron systems considered here, the Hartree-Fock method becomes equivalent to KS DFT with the exact exchange functional, as the non-local HF exchange potential reduces to a local one-body potential. For systems with higher electron number the non-local Hartree-Fock exchange can be transformed into a local potential via the optimized effective potential method, yielding a well defined orbital-dependent functional (called exact exchange). Hybrid functionals are obtained by mixing a fraction of single determinant exchange with GGA or metaGGA functionals, and are normally computed using a non-local potential, a treatment outside the KS framework.

\subsection{Restricted KS DFT with the SCE functional}
\label{sec:SCE}
The HK functional of Eq.~\eqref{eq:UF} and the KS kinetic energy functional of Eq.~\eqref{eq:Ts} can be seen as the values at $\lambda=1$ and $\lambda=0$ of a more general functional $F_\lambda[\rho]$, in which the electronic interaction is rescaled by a coupling strength parameter $\lambda$,
\begin{align}
		F_\lambda[\rho]=\min_{\Psi\rightarrow\rho}\left\langle\Psi | \hat{T} + \lambda\hat{V}_{ee}| \Psi \right\rangle.
		\label{eq:Flambda}
	\end{align}
The SCE functional, first introduced in the seminal work of Seidl and coworkers,\cite{Sei-PRA-99,SeiPerLev-PRA-99} is the strong-interaction limit, $\lambda\rightarrow\infty$, of $F_\lambda[\rho]$, which corresponds to minimizing the electron-electron repulsion alone for a given density $\rho$,
	\begin{align}
		V_{ee}^{\rm SCE}[\rho]=\min_{\Psi\rightarrow\rho}\left\langle\Psi |\hat{V}_{ee}| \Psi \right\rangle.
		\label{eq:VeeSCEdef}
	\end{align}
It is the natural counterpart of the KS kinetic energy functional of Eq.~\eqref{eq:Ts}. The functional $V_{ee}^{\rm SCE}[\rho]$ describes the physical situation in which the electrons are perfectly correlated, so that the position $\rv$ of one of them fixes all the other positions (or, to be more precise, all the interparticle distances) via the so-called {\it co-motion functions} $\textbf{f}_i(\rv)$, $\rv_i=\textbf{f}_i(\rv)$.\cite{SeiGorSav-PRA-07} The co-motion functions are highly non-local functionals of the density $\rho$, satisfying the differential equation\cite{SeiGorSav-PRA-07}
    	\begin{align}
    		\rho(\textbf{f}_i\brv)\td \textbf{f}_i\brv = \rho\brv \td\rv,
    		\label{eq:difff}
    	\end{align}
which can be derived from the constraint ``$\Psi\to\rho$'' of Eq.~\eqref{eq:VeeSCEdef}.\cite{SeiGorSav-PRA-07,GorVigSei-JCTC-09}
They also obey group properties that ensure the indistinguishability of the $N$ electrons,
\begin{equation}
\label{eq_groupprop}
\begin{split}
&\fv_1(\rv) \equiv \rv, \\
&\fv_2(\rv) \equiv \fv(\rv), \\
&\fv_3(\rv) =      \fv(\fv(\rv)), \\
&\fv_4(\rv) =      \fv(\fv(\fv(\rv))), \\
&\qquad\ \vdots \\
&\underbrace{\fv(\fv(\ldots\fv(\fv(\rv))))}_\text{$N$ times} = \rv.
\end{split}
\end{equation}
The minimizing $N$-electron density $|\Psi(\rv_1,...\rv_N)|^2$ in Eq.~\eqref{eq:VeeSCEdef}, which becomes a distribution in this limit,\cite{SeiGorSav-PRA-07,GorVigSei-JCTC-09,ButDepGor-PRA-12,CotFriKlu-CPAM-13} is the strictly-correlated state:
\begin{multline}
\label{eq_psi2SCE}
|\Psi_{\rm SCE}(\rv_1,\rv_2,\dots,\rv_N)|^2 = \frac{1}{N!} \sum_{\wp}
\int d\rv \, \frac{\rho(\rv)}{N} \, \delta(\rv_1-\fv_{\wp(1)}(\rv)) \\
\times\delta(\rv_2-\fv_{\wp(2)}(\rv)) \cdots \delta(\rv_N-\fv_{\wp(N)}(\rv))\; ,
\end{multline}
where $\wp$ denotes a permutation of ${1,\dots,N}$. Equations~\eqref{eq:difff}-\eqref{eq_groupprop} together with the properties of the Dirac $\delta$-function guarantee that 
$\rho(\rv) = N \int |\Psi_{\rm SCE}(\rv,\rv_2,\dots,\rv_N)|^2 
\,d\rv_2\cdots d\rv_N$.
In terms of the co-motion functions, the SCE functional is\cite{SeiGorSav-PRA-07,MirSeiGor-JCTC-12}
	\begin{align}
		V_{ee}^{\rm SCE}[\rho]=\frac{1}{2}\int\td^3r\;\rho\brv\sum_{i=2}^N\frac{1}{|\rv-\textbf{f}_i\brv|},
		\label{eq:VeeSCEfromf}
	\end{align}
and its functional derivative
\begin{equation}
    v_{\rm SCE}(\rv)=\frac{\delta V_{ee}^{\rm SCE}[\rho]}{\delta \rho(\rv)}
\end{equation}
can be obtained from the equation\cite{MalGor-PRL-12,MalMirCreReiGor-PRB-13}
\begin{align}
	\nabla v_{\rm SCE}\brv =- \sum_{i=2}^N\frac{\rv-\textbf{f}_i\brv}{|\rv-\textbf{f}_i\brv|^3},
	\label{eq:vSCE}
\end{align}
which has a simple physical meaning: as the position $\rv$ of one electron fixes all the relative distances, the net electron-electron repulsion acting on an electron at $\rv$ becomes a function of $\rv$ alone, and can be represented as the gradient of a one-body potential. Equation~\eqref{eq:vSCE} is a very powerful shortcut to compute the functional derivative of the highly non-local functional $V_{ee}^{\rm SCE}[\rho]$ of Eqs.~\eqref{eq:difff}-\eqref{eq:VeeSCEfromf}.

The KS SCE approach consists of approximating the constrained minimization in the universal functional $F[\rho]$ of Eq.~\eqref{eq:UF} as the sum of two constrained minima
	\begin{align}
		F[\rho]&\approx \min_{\Psi\rightarrow\rho}\left\langle\Psi | \hat{T}| \Psi \right\rangle + \min_{\Psi\rightarrow\rho}\left\langle\Psi |\hat{V}_{ee}| \Psi \right\rangle\notag\\
		&=T_s[\rho]+V_{ee}^{\rm SCE}[\rho].
		\label{eq:F_KSSCE}
	\end{align}
Obviously, the minimizing wavefunction $\Psi$ is different for $\hat{T}$ and $\hat{V}_{ee}$: for the former, it is usually a single Slater determinant, while for the latter it is the strictly-correlated state of Eq.~\eqref{eq_psi2SCE}. The key point here is that, for a given $\rho$, $V_{ee}^{\rm SCE}[\rho]$ is a well defined {\em density functional}, whose functional derivative can be easily computed via Eq.~\eqref{eq:VeeSCEfromf}. Our total energy functional is then
\begin{equation}
	E[\rho]=T_s[\rho]+V_{ee}^{\rm SCE}[\rho]+\int\rho(\rv)v_{ext}(\rv)\,d\rv.
	\label{eq:totE}
\end{equation}
By varying $E[\rho]$ with respect to the single-particle orbitals appearing in $T_s[\rho]$, one obtains the usual KS equations, so that Eq.~\eqref{eq:F_KSSCE} is completely equivalent to making the approximation
\begin{equation}
    E_{xc}[\rho]\approx V_{ee}^{\rm SCE}[\rho]-U[\rho],
\end{equation}
and thus $v_{xc}(\rv)\approx v_{\rm SCE}(\rv)-u_H(\rv)$.
	\begin{figure}
		\begin{center}
		\includegraphics[width=.48\linewidth]{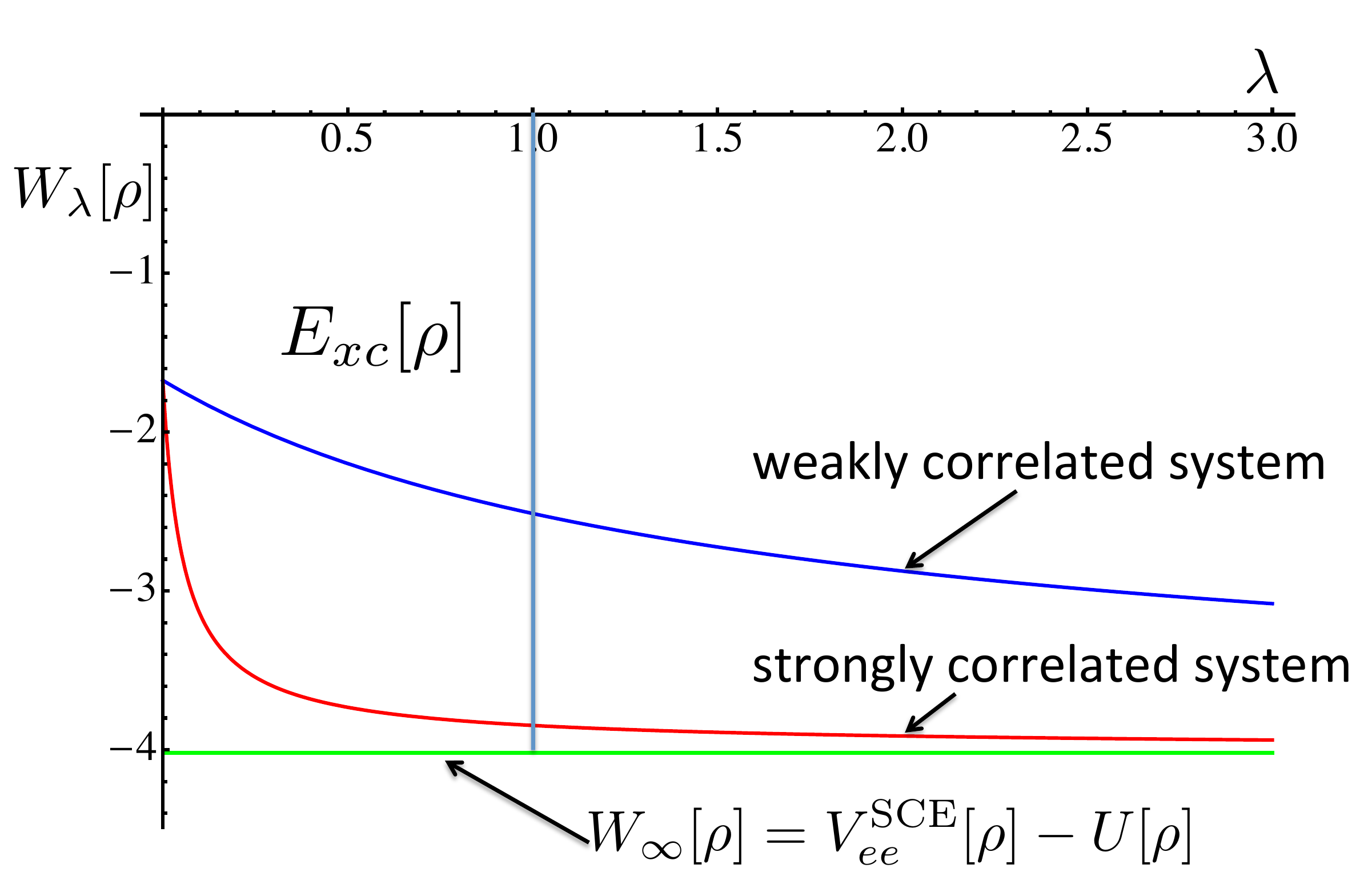}
		\end{center}
		\caption{The KS SCE approximation from the point of view of the adiabatic connection of Eq.~\eqref{eq:adiab}. We have reported $W_\lambda[\rho]$ as a function of $\lambda$ for a typical weakly-correlated and a typical strongly-correlated system. The area between $W_\lambda[\rho]$, the $\lambda$-axis and the vertical lines corresponding to $\lambda=0$ and $\lambda=1$ gives the exchange-correlation energy $E_{xc}[\rho]$. The KS SCE approximates $W_\lambda[\rho]$  with its value at $\lambda\to\infty$ for all $\lambda$, and thus the xc energy with the area of the rectangle limited by $W_\infty[\rho]$, the $\lambda$-axis, and the  vertical lines corresponding to $\lambda=0$ and $\lambda=1$.}
		\label{fig:AdiabConn}
	\end{figure}
The minimization of our energy density functional of Eq.~\eqref{eq:totE} reduces then to solving the standard restricted KS equations self-consistently:
	\begin{align}
		\left[-\frac{1}{2}\nabla^2+v_{\rm SCE}[\rho]\brv+v_{ext}\brv\right]\psi_i\brv=\epsilon_i\, \psi_i\brv.
	\end{align}	
Notice that the self-consistent KS SCE total energy that we obtain in this way is always a lower bound to the exact one. In fact, since the minimum of a sum is always larger or equal than the sum of the minima, for the exact ground-state density $\rho$ we have
\begin{align}
	F[\rho]+\int\rho\; v_{ext} \geq T_s[\rho]+V_{ee}^{\rm SCE}[\rho] +\int\rho\; v_{ext}.
\end{align}
This inequality becomes even stronger when we minimize the right-hand side by solving self-consistently the KS SCE equations. The SCE functional is also self-interaction free, as $V_{ee}^{\rm SCE}[\rho]=0$ for any one-electron density. Thus, as $E_{N=2}^{\rm KS\; SCE}\le E_{N=2}^{\rm exact}$ and $E_{N=1}^{\rm KS \;SCE}= E_{N=1}^{\rm exact}$, the self-consistent KS SCE method will certainly bind all the anions of the He isoelectronic series that are physically bound, and its error will always be towards overbinding, providing a lower bound for $Z_{\rm crit}$, which, however, turns out to be not very tight (see Sec.~\ref{sec:results}).

It is also useful to represent graphically the approximation made in KS SCE in terms of the standard adiabatic connection of KS DFT\cite{LanPer-SSC-75} (see Fig.~\ref{fig:AdiabConn}). By denoting $\Psi_\lambda[\rho]$ the minimizing wave function in Eq.~\eqref{eq:Flambda}, and by defining the indirect part $W_{\lambda}[\rho]$ of the electron-electron repulsion at coupling strength $\lambda$,
\begin{equation}
    W_{\lambda}[\rho]=\langle\Psi_\lambda[\rho]|\hat{V}_{ee}|\Psi_\lambda[\rho]\rangle-U[\rho],
\end{equation}
one obtains the well-known exact formula\cite{LanPer-SSC-75} for $E_{xc}[\rho]$,
\begin{equation}
    E_{xc}[\rho]=\int_0^1 W_{\lambda}[\rho]\, d\lambda.
    \label{eq:adiab}
\end{equation}
In Fig.~\ref{fig:AdiabConn} we show, schematically, $W_{\lambda}[\rho]$ as a function of $\lambda$ for a weakly- and a strongly-correlated system. The area between $W_\lambda[\rho]$, the $\lambda$-axis and the vertical lines corresponding to $\lambda=0$ and $\lambda=1$ gives the exchange-correlation energy $E_{xc}[\rho]$. The KS SCE approximates $W_\lambda[\rho]$  with its value at $\lambda\to\infty$ for all $\lambda$, and thus the \xc energy with the area of the rectangle limited by $W_\infty[\rho]$, the $\lambda$-axis, and the  vertical lines corresponding to $\lambda=0$ and $\lambda=1$. This is evidently a good approximation only when the system is very correlated.

Evaluating the co-motion functions in the general case is still an open problem, although progress has been made\cite{MenLin-PRB-13} by using the dual Kantorovich formulation,\cite{ButDepGor-PRA-12} which allows one to evaluate $V_{ee}^{\rm SCE}[\rho]$ and its functional derivative $v_{\rm SCE}(\rv)$ in a different way, bypassing the co-motion functions. In the special case of spherically symmetric densities, like the ones considered here, an explicit solution is known\cite{SeiGorSav-PRA-07} in terms of the function $N_e(r)$,
\begin{equation}
    N_e(r)=\int_0^r 4\pi x^2\rho(x)\, dx
    \label{eq:Ne}
\end{equation}
and its inverse $N_e^{-1}$. For a $N=2$ system, the two electrons in the SCE solution are always opposite to each other with respect to the nucleus (maximum angular correlation), at a relative angle $\pi$. Their distances from the nucleus, $r_1=r$ and $r_2=f(r)$, are related by the single co-motion function
\begin{equation}
    f(r)=N_e^{-1}[2-N_e(r)].
    \label{eq:f}
\end{equation}
Equations~\eqref{eq:Ne}-\eqref{eq:f} clearly show the non-local dependence of $f(r)$ on the density. The SCE potential $v_{\rm SCE}(r)$ is then simply obtained by integrating the spherically-symmetric equivalent of Eq.~\eqref{eq:vSCE},
\begin{equation}
    v_{\rm SCE}'(r)=-\frac{1}{[r+f(r)]^2},
    \label{eq:vSCEspher}
\end{equation}
with boundary condition $v_{\rm SCE}(r\to\infty)=0$. Notice that $v_{\rm SCE}(r)$ has the correct asymptotic behavior of the Hartree plus xc potential, $v_{\rm SCE}(r \to\infty)\sim 1/r$, since $f(r\to \infty)\to 0$. This is true for the general $N$-electron case also, since the correct $(N-1)/r$ asymptotic leading term can be similarly derived\cite{SeiGorSav-PRA-07} from Eq.~\eqref{eq:vSCE}.

\subsection{Local corrections to the SCE functional}
\label{sec:KSSCELDA}
As can be expected from Fig.~\ref{fig:AdiabConn}, KS SCE can capture correlation effects at all correlation regimes, but good quantitative accuracy is obtained only when correlation becomes very strong.\cite{MalGor-PRL-12,MalMirCreReiGor-PRB-13,MenMalGor-PRB-14} In practice, however, the systems of interest in chemistry are in between the weak- and strong correlation regimes and it is thus desirable to improve the approximation of Eq.~\eqref{eq:F_KSSCE}. Therefore, we consider the more general decomposition of the universal functional
	\begin{align}
		F[\rho]=T_s[\rho]+V_{ee}^{\rm SCE}[\rho]+T_c[\rho]+V_{ee}^d[\rho],
		\label{eq:F_SCE+LDA}
	\end{align}
where $T_c[\rho]$ (kinetic correlation energy) is the difference between the true kinetic energy and the KS one,
	\begin{align}
		T_c[\rho]=\left\langle\Psi_{\lambda=1}[\rho] |\hat{T}| \Psi_{\lambda=1}[\rho] \right\rangle - T_s[\rho],
	\end{align}
and $V_{ee}^d[\rho]$ (decorrelation energy\cite{GorSeiVig-PRL-09,LiuBur-JCP-09}) is the difference between the true electron-electron repulsion energy  and the SCE value,
	\begin{align}
		V_{ee}^d[\rho]=\left\langle\Psi_{\lambda=1}[\rho] |\hat{V}_{ee}| \Psi_{\lambda=1}[\rho] \right\rangle - V_{ee}^{\rm SCE}[\rho].
	\end{align}
Both corrections are evidently always positive.
A simple way to construct the correcting term $T_c[\rho]+V_{ee}^d[\rho]$ is to make a local density approximation, which can be defined as the correction that makes Eq.~\eqref{eq:F_SCE+LDA} exact when the density $\rho(\rv)$ becomes uniform,
	\begin{align}
		T_c^{\rm LDA}[\rho]+V_{ee}^{d, {\rm LDA}}[\rho]=\int\td^3r\;\rho\brv\left\{(t_c[\rho\brv]+v_{ee}^d[\rho\brv]\right\},
	\end{align}
where $t_c(\rho)$ and $v_{ee}^d(\rho)$ are the kinetic correlation energy per particle and the electron-electron decorrelation energy per particle of the homogeneous electron gas (HEG) of density $\rho$. They can be easily obtained as
	\begin{align}
		t_c(\rho)+v_{ee}^d(\rho)=\epsilon_{xc}(\rho)-\epsilon_{\rm SCE}(\rho),
		\label{eq:loccorr}
	\end{align}
where $\epsilon_{xc}(\rho)$ and $\epsilon_{\rm SCE}(\rho)$ are, respectively, the \xc energy per particle and the indirect part of the SCE interaction energy per particle for the HEG. The latter can be obtained by considering that in the external potential due to an infinite uniform background with positive charge density $\rho^+=(\frac{4}{3}\pi r_s^3)^{-1}$ the minimum possible electron-electron repulsion is attained with the electrons localized at the sites of the bcc crystal with lattice parameter $a=2(\pi/3)^{1/3} r_s$. A uniform electronic density $\rho=\rho^+$ is constructed by taking a linear superposition of all the possible origins and orientations of the crystal. In other words, in the simple uniform-density case, the co-motion functions are just the lattice vectors of the bcc crystal with origin in the reference electron, whose position is distributed uniformly. This means that {\em for all values of the density parameter $r_s$} the SCE energy of the uniform electron gas is equal to the low-density leading term of the HEG energy,
	\begin{align}
		\epsilon_{\rm SCE}(\rho)=-\frac{d_0}{r_s(\rho)}.
	\end{align}
At high densities, the SCE energy is very far from the exact one, and at low densities it becomes asymptotically exact. This is also true, more generally, for the functional $V_{ee}^{\rm SCE}[\rho]$, which approaches the exact Hartree plus exchange correlation functional when the density is scaled as $\rho_\gamma(\rv) = \gamma^3\rho(\gamma \rv)$ and $\gamma \to 0$.
Here we have set $d_0\approx 0.891687$, which is the value from the Perdew-Wang-92 LDA parametrization.\cite{PerWan-PRB-92} We denote this method KS SCE+LDA.

It is also possible to consider the local correction only for the electron-electron repulsion part, assuming that the error made by the KS kinetic energy is, for these systems, less serious than the one made by the SCE functional, so that the correction needs to rebalance the two terms. This corresponds to taking as correction only
\begin{align}
	V_{ee}^{d, {\rm LDA}}[\rho]=\int\td^3r\;\rho\brv v_{ee}^d[\rho\brv],
\end{align}
where $v_{ee}^d(r_s)$ is obtained by subtracting from Eq.~\eqref{eq:loccorr} the kinetic correlation contribution $t_c=-\frac{d}{d r_s}(r_s \epsilon_{xc})$. We call this approximation KS SCE+L$V_{ee,d}$.

\section{Results}\label{sec:results}

\subsection{Accurate solution and exact properties at $Z_{\rm crit}$}
\begin{figure}
		\begin{center}
		\includegraphics[width=.48\linewidth]{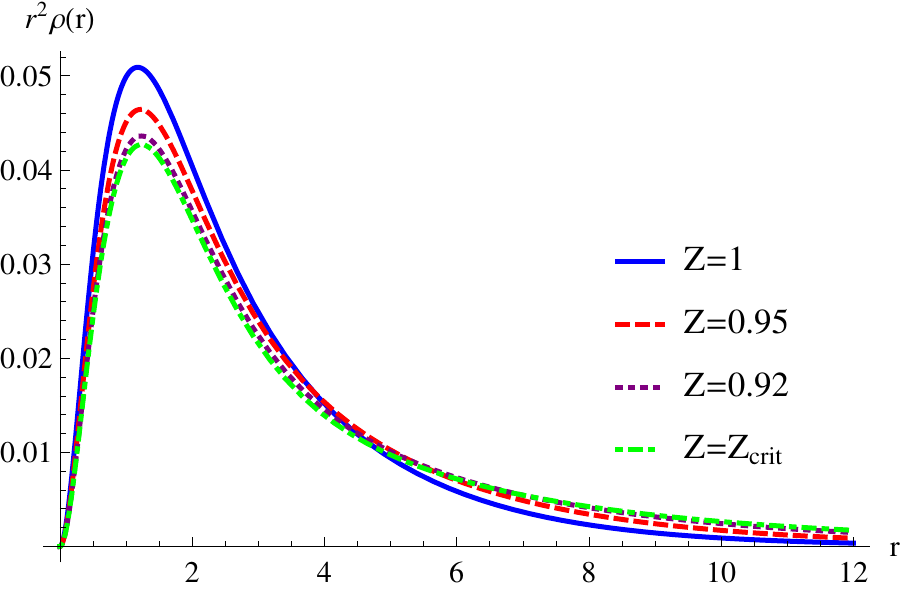}
		\includegraphics[width=.48\linewidth]{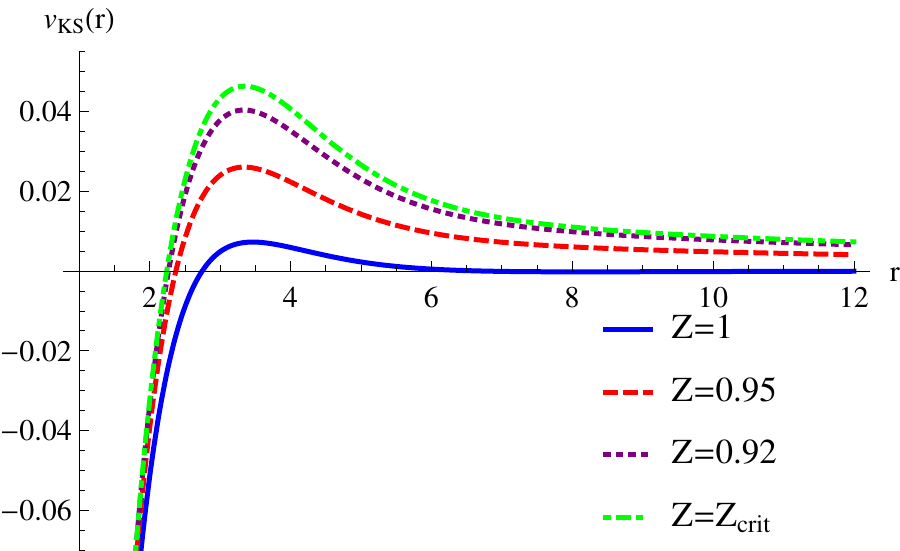}
		\end{center}
		\caption{Accurate $r^2\rho(r)$ and $v_{\rm KS}(r)$ for various anions with nuclear charge $Z$ of the He isoelectronic series.}
		\label{fig:Zcrit_MU}
	\end{figure}

Before presenting and discussing the KS SCE results, we extend the work of Umrigar and Gonze\cite{UmrGon-PRA-94} by studying the accurate densities and KS potentials obtained from the wavefunctions of Sec.~\ref{sec:wavef} close to the quantum phase transition. The densities and KS potentials (obtained by inversion of the KS equations)\cite{UmrGon-PRA-94} for selected values of $Z\leq 1$ are shown in Fig.~\ref{fig:Zcrit_MU}.

The exact density of an atomic or molecular system is known to decay (with exceptions when the ground-state of the ion is not asymptotically accessible by symmetry) as\cite{KatDav-PNAS-80,LevPerSah-PRA-84,AlmBar-PRB-85} $\rho(r\rightarrow\infty)\sim\exp(-2\sqrt{2\,I_p}\,r)$, an expansion which is valid for $1/r \ll I_p$, where $I_p$ is the ionization energy. When $Z\to Z_{\rm crit}$ ($I_p \to 0$) the density remains compact, in agreement with the rigorous result of Ref.~\onlinecite{HofHofSim-JPA-83}, where it has been proven that the density at $Z_{\rm crit}$ satisfies
	\begin{align}
		C_-(\delta) r^{-3/2-\delta}e^{-2\left[8\left(1-Z_{\rm crit}\right)r\right]^{1/2}} \leq \rho(r) \leq C_+(\delta) r^{-3/2+\delta}e^{-2\left[8\left(1-Z_{\rm crit}\right)r\right]^{1/2}},
		\label{eq:dec_Zcrit}
	\end{align}
where $\delta$ is an arbitrary small positive number and $C_\pm(\delta)$ are constants depending on $\delta$.

\begin{figure}
		\begin{center}
		\includegraphics[width=.8\linewidth]{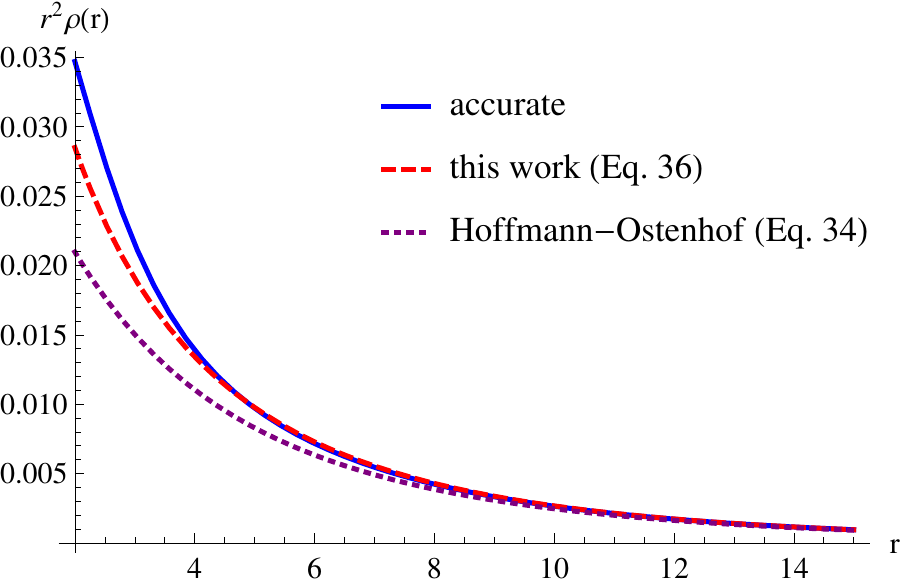}
		\end{center}
		\caption{Comparison of the long-range behaviour of $r^2\rho(r)$ at $Z_{\rm crit}$ obtained from the asymptotic decay expressions in Eqs.~\eqref{eq:dec_Zcrit} and \eqref{eq:dec_Zcrit2} with the almost exact result obtained from the wavefunction in Section~\ref{sec:wavef}.  }
		\label{fig:Zcrit_asympt}
	\end{figure}

We can further understand the asymptotic decay of the density at $Z_{\rm crit}$ by studying the corresponding differential equation\cite{LevPerSah-PRA-84} for $\sqrt{\rho}$ (which for a $N=2$ singlet coincides with the KS equation). At the quantum phase transition with the asymptotic potential to fourth order\cite{AlmBar-PRB-85,UmrGon-PRA-94} this equation is
	\begin{align}
		\left[-\frac{1}{2}\nabla_r^2 - \frac{Z-N+1}{r} + O\left(\frac{1}{r^4}\right)\right]\sqrt{\rho(r)}=0.
		\label{eq:asympt}
	\end{align}
By solving Eq.~\eqref{eq:asympt} asymptotically ($r\to\infty$), we obtain, order by order, a solution for the leading terms to order $O(r^{-4})$,
	\begin{align}
		\rho(r\rightarrow \infty)\sim &\; \frac{e^{-4a\sqrt{r}}}{r^{3/2}}\left(1+\frac{3}{8a\,r^{1/2}}-\frac{3}{128a^2\,r}+\frac{15}{1024a^3\,r^{3/2}}-\frac{405}{32768a^4\,r^2}\right),
		\label{eq:dec_Zcrit2}
	\end{align}
with $a=\sqrt{2(-Z+N-1)}$. This decay agrees to leading order with Eq.~\eqref{eq:dec_Zcrit}.
The accurate density at the quantum phase transition together with the decays from Eqs.~\eqref{eq:dec_Zcrit} and \eqref{eq:dec_Zcrit2} are displayed in Fig.~\ref{fig:Zcrit_asympt}, where in both cases the proportionality constant has been adjusted to match the accurate density at the end of the radial grid ($r\approx 100$). Notice that Eq.~\eqref{eq:asympt} implies that for the {\em exact} KS system (which yields the {\em exact} ground-state density) the equality $\epsilon_{\rm HOMO}=-I_p$ also holds at $Z=Z_{\rm crit}$, when $I_p=0$.

From Fig.~\ref{fig:Zcrit_MU} we see that the KS potentials have a bump at intermediate length scale. This bump increases for smaller $Z$ as can be expected from the asymptotic first order contribution at large $r$,  $v_{\rm KS}(r\to\infty)= (1-Z)/r$ that will be positive for $Z<1$. The bump is present also for the Hydrogen anion, where this first order contribution vanishes.

\begin{figure}
		\begin{center}
		\includegraphics[width=.8\linewidth]{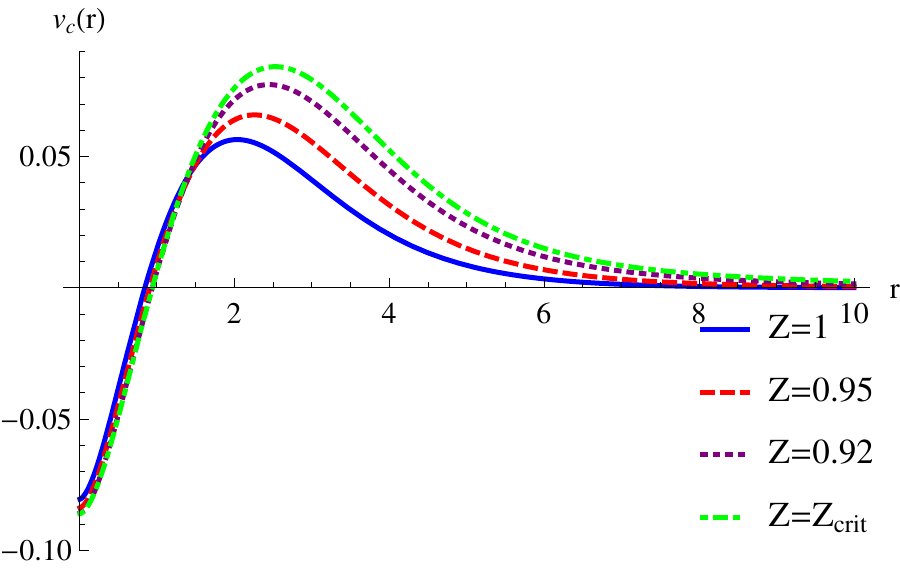}
		\end{center}
		\caption{Accurate correlation potential $v_c(r)$ for various $Z$.}
		\label{fig:vc_MU}
	\end{figure}

In Fig.~\ref{fig:vc_MU} we show the correlation potentials for selected values of $Z$. 
We see that, as was found in Ref.~\onlinecite{UmrGon-PRA-94}, the accurate correlation potential close to the nucleus has a nearly quadratic behavior. In Refs.~\onlinecite{Qia-PRB-07,QiaSah-PRA-07} it has been shown that the linear term in the correlation potential is due to the kinetic contribution, which, thus, turns out to be very small.
	

\subsection{$Z_{\rm crit}$ from KS-DFT with standard functionals and from KS SCE}
For the He isoelectronic series with $Z\leq 2$ we solved self-consistently the restricted KS equations with various approximate functionals. Calculations for the HF (or exact exchange) method, KS LDA, KS SCE, and KS SCE with the two local corrections of Sec.~\ref{sec:KSSCELDA} were performed with a numerical code developed in our group. We chose the Perdew-Wang-92 functional (PW92)\cite{PerWan-PRB-92} LDA parametrization. To compare our calculations with the available standard approximations we have further performed restricted KS-DFT calculations with the Amsterdam Density Functional package (ADF).\cite{ADF1, *ADF2, *ADF3} From the GGA class of functionals we chose PBE,\cite{PerBurErn-PRL-96} from the metaGGA class the revTPSS\cite{TaoPerStaScu-PRL-03} functional and for the hybrid functional we chose B3LYP.\cite{Bec-JCP-93, SteDevChaFri-JPC-94, LeeYanPar-PRB-88, VosWilNus-CJP-80} If not mentioned otherwise, all ADF calculations were carried out in the even-tempered (ET) QZ3P basis supported by 3 diffuse s-functions with the parametrization of Hydrogen.\cite{ChoLenGisBae-JCC-04} To assess the quality of the basis set we also performed KS-LDA (PW92 functional) calculations with the ADF package and compare them to our numerical solution of the KS equations.
To assess the quality of the basis set we also performed KS-LDA calculations with the ADF package (PW92 functional) and compare them to our numerical solution of the KS equations.

We define the critical nuclear charge $Z_{\rm crit}$ for the various DFT approximations to be the value of $Z$ at which either the ionization energy $I_p=E_{N-1}-E_N$ becomes smaller than $0$ or the HOMO eigenvalue $\epsilon_{\rm HOMO}$ becomes positive, whichever is larger.
Although the equality $\epsilon_{\rm HOMO}=-I_p$ does not hold in general for approximate functionals, we invoke the HOMO eigenvalue criterion to avoid the conceptual and numerical issue of occupying orbitals with a positive eigenvalue already discussed.

Table \ref{tab:Zcrit} shows the predicted $Z_{\rm crit}$ for the quantum phase transition together with the corresponding ionization energy $I_p=E_{N=1}-E_{N=2}$ and the HOMO energies for the various approximations. Of the DFT approximations considered only the SCE functionals (SCE and SCE with local corrections) and the hybrid functional are able to bind the Hydrogen anion. The hybrid functional however, yields an unphysical description of the bound anion as we will further discuss below. Remarkably, all the standard functionals at different levels of approximation yield a similar value of $Z_{\rm crit}\approx 1.2$. This shows that the nonlocality encoded in the SCE functional is able to capture different many-body effects than the standard approximations.

\begin{table}
       \caption{$Z_{crit}$ with corresponding negative ionization energy $-I_p=E_{N=2}-E_{N=1}$ and HOMO energies for various approaches.}
       \centering
       \begin{tabular}{l|cccc}
               \hline
                   & $Z_{crit}$ & $\epsilon_{\rm HOMO}$ & $-I_p$\\
               \hline
               Accurate & 0.9110 & 0.0 & 0.0\\
               \hline
               \emph{numerical}\\
               \quad HF  & 1.0312 & -0.05809 & 0.0\\
               \quad KS-LDA, PW92  & 1.2244 & 0.0 & -0.18509\\
               \quad KS-SCE  & 0.7307 & 0.0 & -0.05639 \\
               \quad KS-SCE+LDA, PW92 & 0.9474 & 0.0 & -0.05253\\
               \quad KS-SCE+LVee,d, PW92 & 0.9012 & 0.0 & -0.04964\\
               \hline
               \emph{ET-QZ3P+3diffuse}\\
               \quad KS-LDA, PW92 & 1.2240 & 0.0 & -0.18477\\
               \quad KS-GGA, PBE & 1.2303 & 0.0 & -0.19179\\
               \quad KS-metaGGA, revTPSS & 1.2120 & 0.0 & -\footnote{Not supported by ADF\cite{ADF1, *ADF2, *ADF3}}\\
               \quad KS-Hybrid, B3LYP & 0.6932 & -0.0041 & 0.0\\
               \quad (KS-Hybrid, B3LYP)\footnote{ET-QZ3P basis} & 1.1403 & 0.0 & -0.15909\\
               \hline
       \end{tabular}
       \label{tab:Zcrit}
   \end{table}
		
\begin{figure}
		\begin{center}
		\includegraphics[width=.8\linewidth]{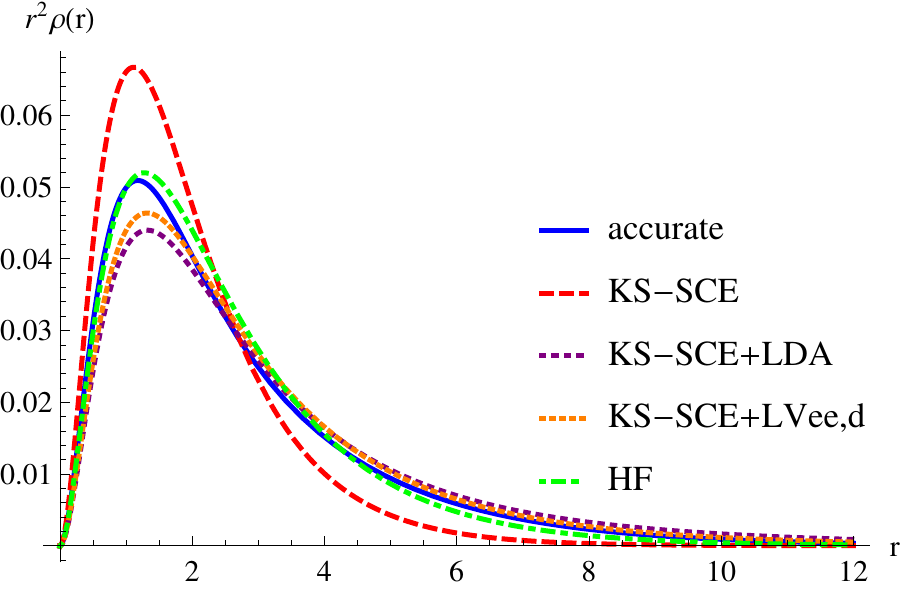}
		\end{center}
		\caption{$r^2\rho(r)$ for H$^-$ and various approaches.}
		\label{fig:dens_H-}
	\end{figure}

As already discussed in Sec.~\ref{sec:SCE}, the KS SCE self-consistent results yield a lower bound to the total energy, which for these systems is not very tight as can be seen from the underestimation of $Z_{\rm crit}\approx 0.7307$ versus the actual value $Z_{\rm crit}\approx 0.9110289$. This is due to the inherent strong correlation nature of the electrons in the SCE formulation that results in underestimating the electron-electron repulsion energy. Self-consistently thus, the KS-SCE densities become quite compact until the kinetic energy starts to dominate in Eq.~\eqref{eq:F_KSSCE}. This is manifested in Fig.~\ref{fig:dens_H-} where the density of H$^-$ is displayed for several methods, the KS-SCE yielding the most compact density. Physically, this is due to the fact that the two electrons, being perfectly correlated, can avoid each other as much as possible and can get much closer to the nucleus to lower the total energy.

	\begin{figure}
		\begin{center}
		\includegraphics[width=.48\linewidth]{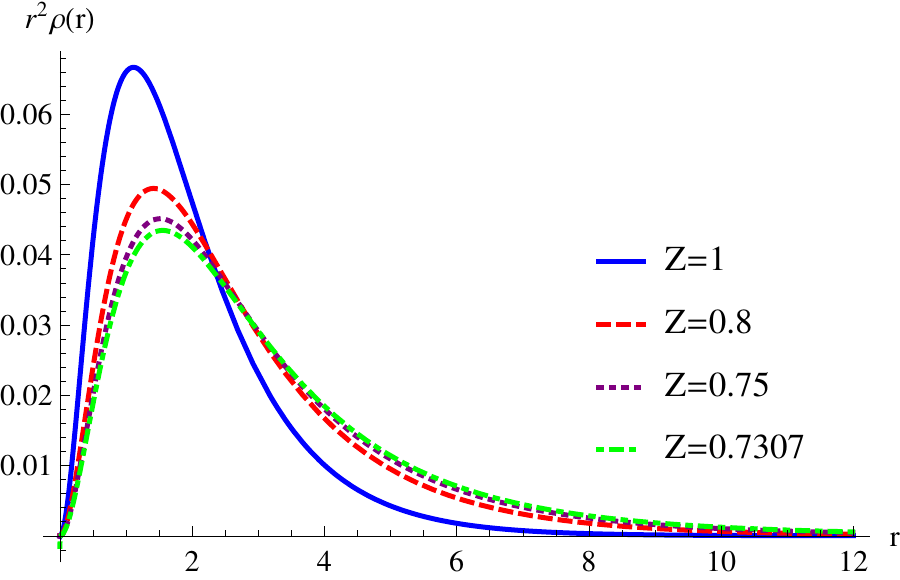}
		\includegraphics[width=.48\linewidth]{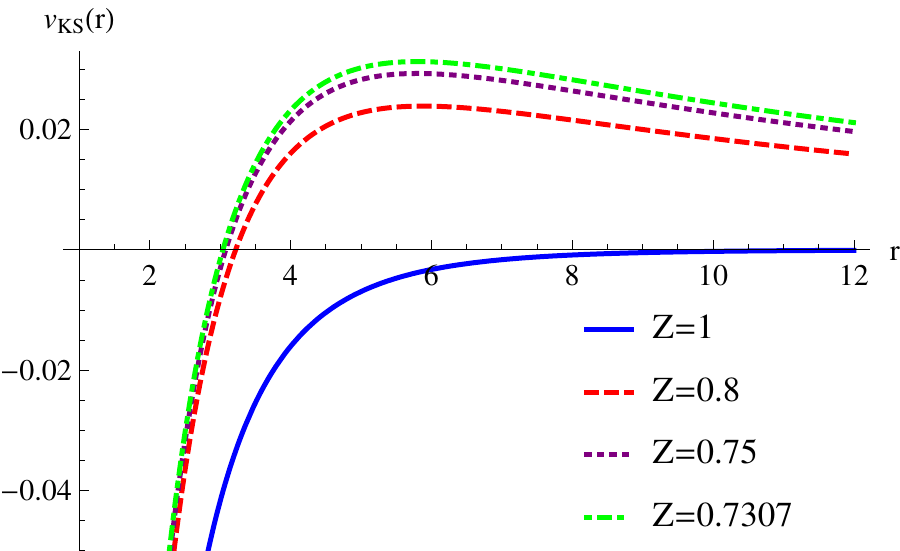}
		\includegraphics[width=.48\linewidth]{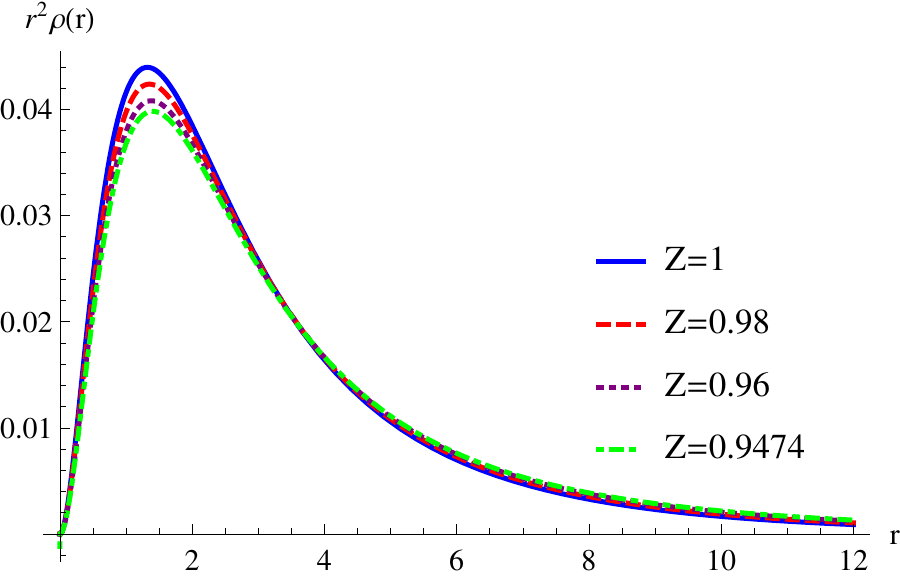}
		\includegraphics[width=.48\linewidth]{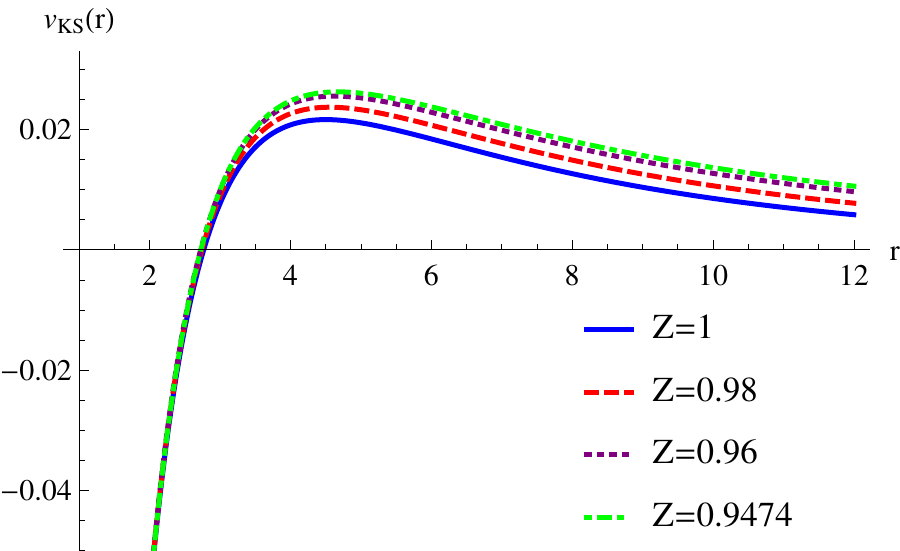}
		\end{center}
		\caption{$r^2\rho(r)$ and $v_{\rm KS}$ for various $Z$ for the KS-SCE (above) and KS-SCE+LDA (below) method.}
		\label{fig:Zcrit_SCEs}
	\end{figure}

The local corrections to the SCE functional improve considerably the predicted $Z_{\rm crit}$ and give a more realistic description of the electronic interactions. We observe that the KS-SCE+LDA density is too spread out compared to the accurate data (e.g. Fig.~\ref{fig:dens_H-}). This can be attributed to the self-interaction error that is introduced by the LDA correction which is obvious from Eq.~\eqref{eq:loccorr} -- the energy densities do not vanish for a density integrating to 1.

In Fig.~\ref{fig:dens_H-} we display also the Hartree-Fock density. It is possible to do this because the HOMO eigenvalue
is negative even though $E_{N=1}<E_{N=2}$.  (If the electron number were treated as a variational parameter, the minimum
energy would be attained for $N<2$.)  We see that the HF density resembles the accurate density more closely than the density from other functionals considered.  This supports the point of view of Ref.~\onlinecite{KimSimBur-PRL-13}, and the general idea
of using HF densities as input for DFT energies in the case of negative ions,\cite{LeeFurBur-JPCL-10,KimSimBur-JCP-11}
even when HF does not bind the last electron.

\begin{figure}
		\begin{center}
		\includegraphics[width=.8\linewidth]{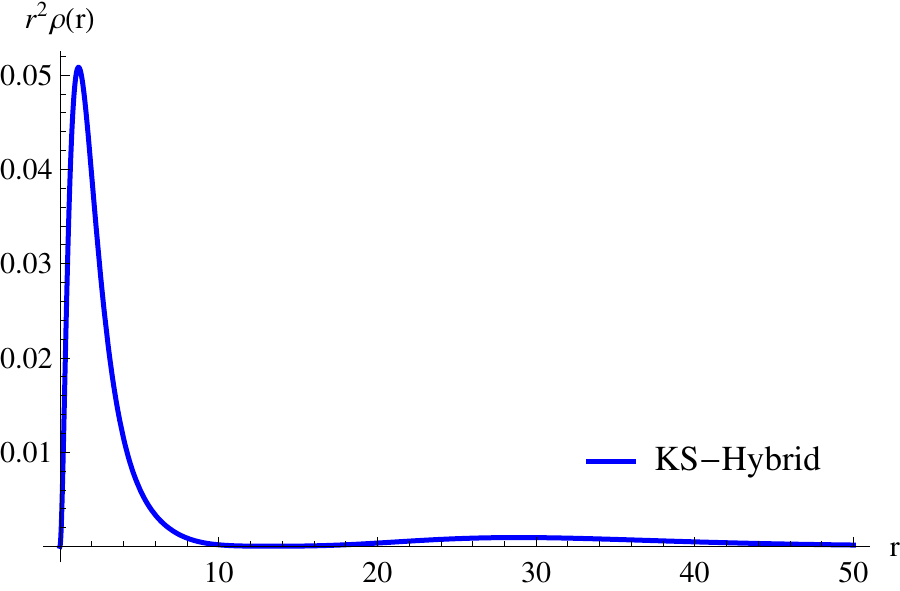}
		\end{center}
		\caption{$r^2\rho(r)$ for H$^-$ with the B3LYP functional and a ET-QZ3P plus 3 diffuse function basis. It displays a second unphysical maximum in the density.}
		\label{fig:dens_H-_hyb}
	\end{figure}

For the hybrid functional in the ET-QZ3P+3diffuse basis we obtain a negative $\epsilon_{\rm HOMO}$ and $E_{N=1}>E_{N=2}$ for H$^-$. Formally the hybrid thus binds the Hydrogen anion. When inspecting the density however, one
observes that it escapes partially from the nucleus, as shown in Fig.~\ref{fig:dens_H-_hyb}.
When removing the 3 diffuse basis functions from the basis set to prevent the density accumulation in the outside regions, we obtain a value of $Z_{\rm crit}$ in between that from HF and conventional DFT, as expected.

\begin{figure}
		\begin{center}
		\includegraphics[width=.8\linewidth]{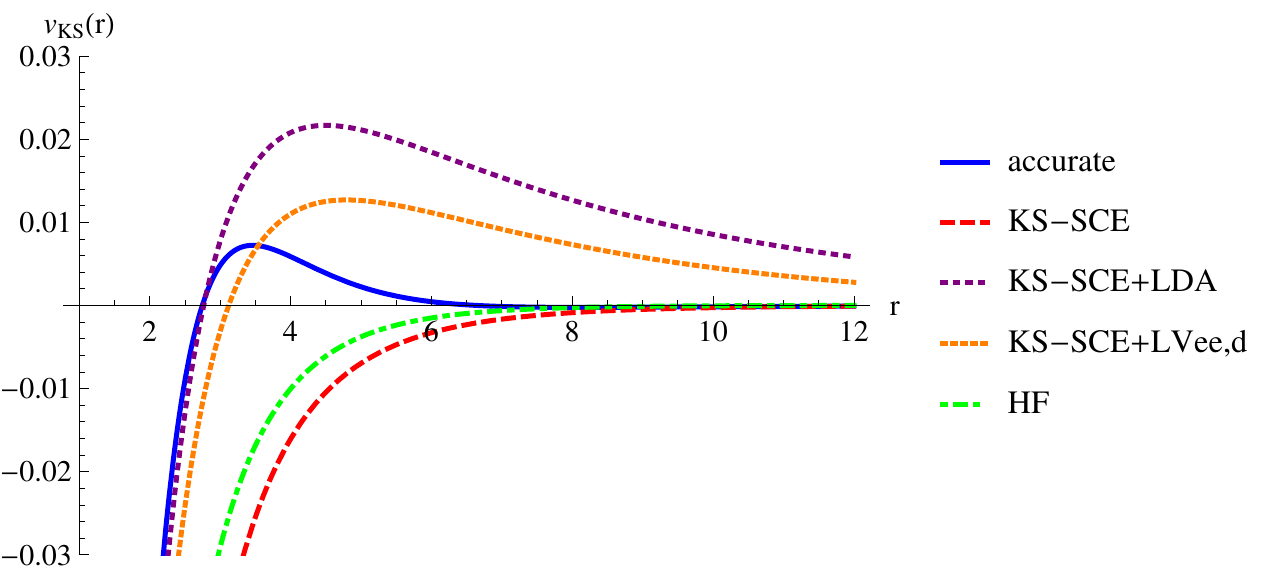}
		\includegraphics[width=.8\linewidth]{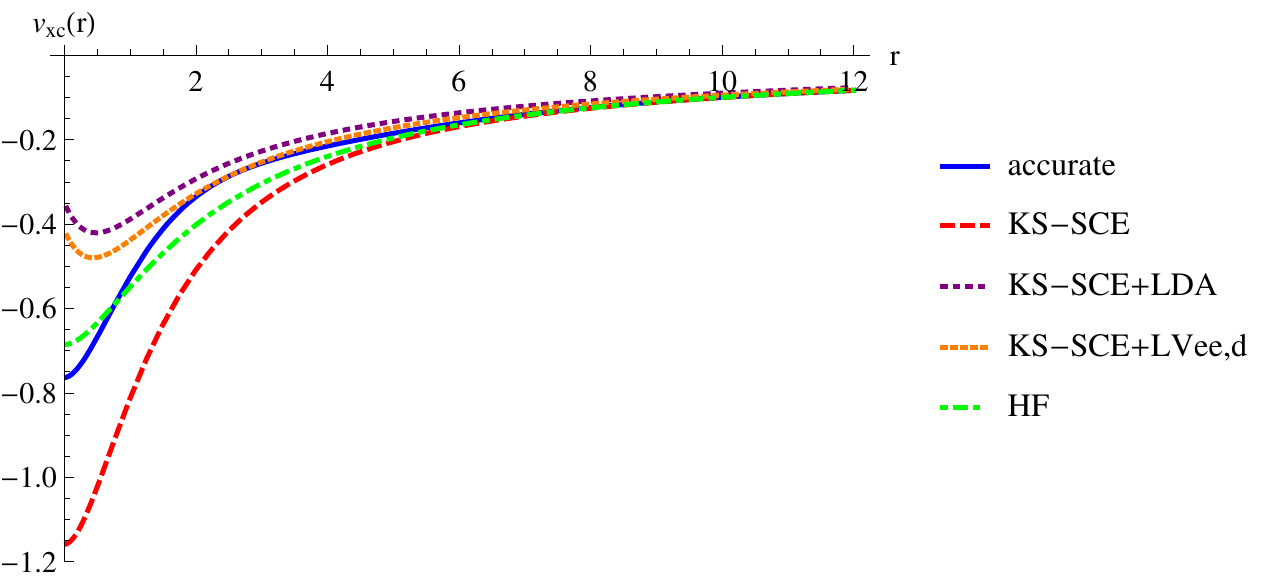}
		\end{center}
		\caption{$v_{\rm KS}(r)$ and $v_{xc}(r)$ for H$^-$ and various approaches.}
		\label{fig:vks_H-}
	\end{figure}

We now discuss the Kohn-Sham and \xc potentials for the self-consistent densities, displayed in Figs.~\ref{fig:Zcrit_SCEs} and~\ref{fig:vks_H-}.
We see that the SCE total Kohn-Sham potential does not develop the bump for $Z=1$, but only for smaller nuclear charges when the interelectronic repulsion dominates over the weaker nuclear attraction (Fig.~\ref{fig:Zcrit_SCEs}). As already observed in Fig.~\ref{fig:dens_H-}, this corresponds to a very compact density. For larger distances, the SCE potential is in good agreement with the accurate one, as expected from the absence of the self-interaction error in the SCE. From  Fig.~\ref{fig:vks_H-} we also see that the SCE potential is quadratic close to the nucleus, as can be easily proven analytically from Eq.~\eqref{eq:vSCEspher}, since when $r\to 0$ we have $f(r\to 0)\to\infty$, so that $v'_{\rm SCE}(r\to 0)=0$.  This is in agreement with the findings of Refs.~\onlinecite{Qia-PRB-07,QiaSah-PRA-07}, as there is no kinetic contribution in the SCE potential. 

\begin{figure}
	\begin{center}
	\includegraphics[width=.8\linewidth]{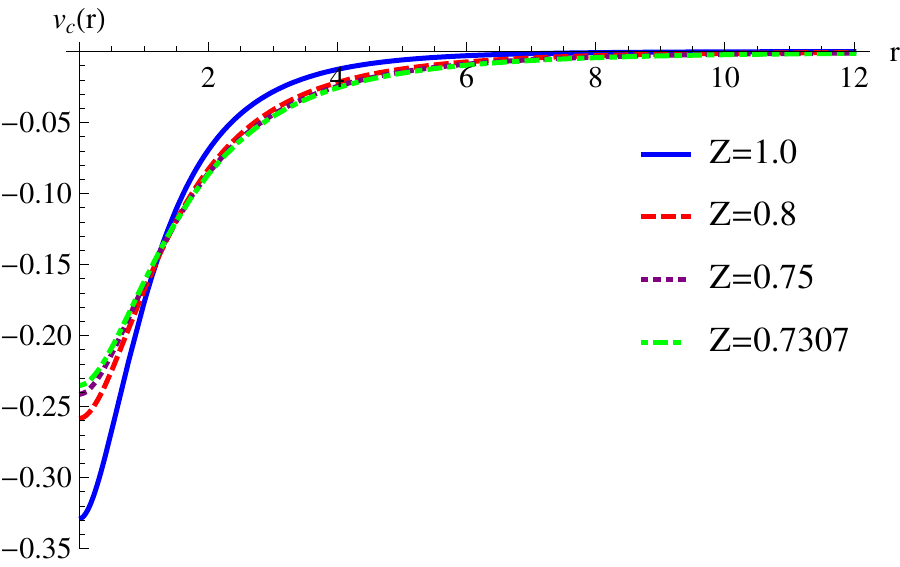}
	\end{center}
	\caption{The self-consistent correlation potentials $v_c(r)$ from the bare KS SCE method.}
	\label{fig:vcSCE}
\end{figure}

Although the SCE functional approximates exchange and correlation together, in Fig.~\ref{fig:vcSCE} we show the SCE correlation potential alone, obtained by subtracting from the xc SCE potential the exchange potential constructed from the self-consistent KS SCE densities. We see that the SCE correlation potential is always negative, in contrast to the exact one. The positive part of the exact correlation potential is mainly due to kinetic correlation effects\cite{BuiBaeSni-PRA-89,GriLeeBae-JCP-96} that are missed in the bare SCE. 

At least qualitatively, the bump  for H$^{-}$ in the total KS potential is captured by the KS-SCE with the two local corrections, though the bump is too pronounced, particularly in KS SCE+LDA. This is also responsible for the overestimation of $Z_{\rm crit}$ and can be partially attributed to the self-interaction error. However, the self-interaction error present in the KS-SCE+LDA approach is substantially different from the self-interaction error in standard KS-LDA or KS-GGA. In KS-LDA and GGA the self-interaction error manifests in the wrong asymptotic decay of the KS potential ($-\frac{Z-N}{r}$ instead of $-\frac{Z-N+1}{r}$). KS-SCE has the correct $-\frac{Z-N+1}{r}$ decay and this is not altered by the exponentially vanishing LDA contribution upon going from KS-SCE to KS-SCE-LDA. The KS SCE+L$V_{ee,d}$ is more attractive at short distance than the exact KS potential, achieving error compensation with the overestimation of the bump (less severe than in the KS-SCE+LDA method), which results in a good estimate for $Z_{\rm crit}\approx 0.9012$.
Of the methods studied, the KS-SCE approach with the local corrections is the one in which the HOMO energy deviates the least from the corresponding $E_{N}-E_{N-1}$ (see Tab.~\ref{tab:Zcrit}).

The HF (or exact exchange) potential is also shown in Fig.~\ref{fig:vks_H-}, although, once more, we have to keep in mind that in this case $E_{N=2}>E_{N=1}$, so that the system is not really physically bound.
	
\subsection{Fractional Electron Numbers at $Z=1$}
We complete our analysis by also allowing for fractional electron numbers $Q$, with $0\le Q \le 2$, in the Hydrogen nuclear potential, which is often considered a paradigmatic model for a Mott insulator.\cite{MorCohYan-PRL-09} In exact KS DFT, it is known that the HOMO eigenvalue should be constant between any two adjacent integer electron numbers (say, $N$ and $N+1$), equal to the negative of the exact, interacting, ionization energy $-I_p=E_{N+1}-E_N$, and should jump whenever an integer electron number is crossed.\cite{PerParLevBal-PRL-82,SagPer-PRA-08}
\begin{figure}
		\begin{center}
		\includegraphics[width=1\linewidth]{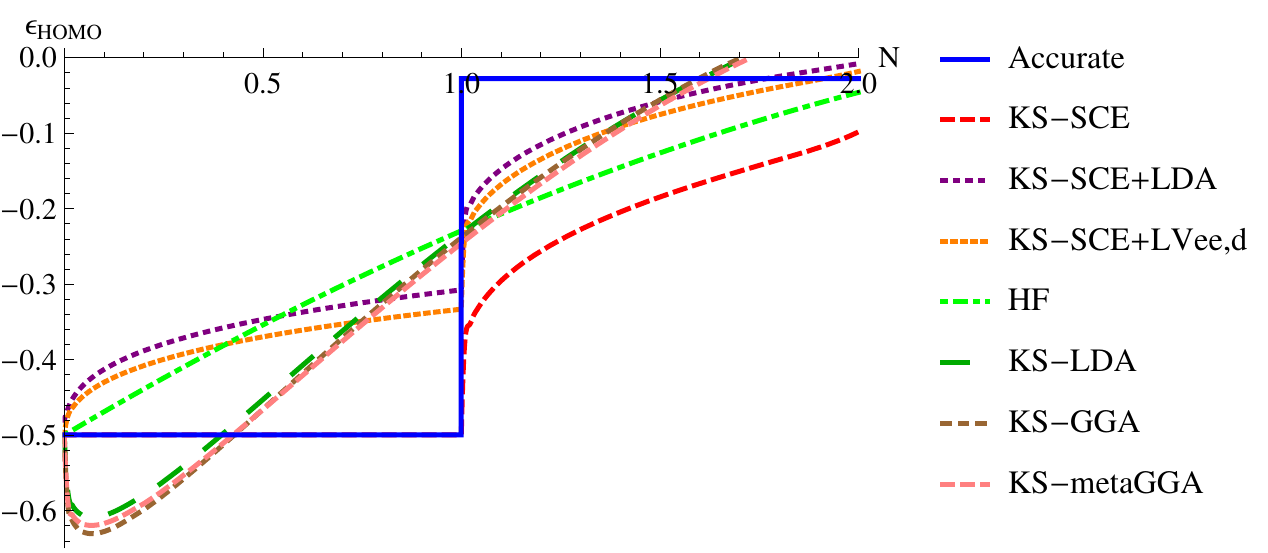}
		\end{center}
		\caption{$\epsilon_{\rm HOMO}$ vs. $N$ in the Hydrogen nuclear potential for various approaches.}
		\label{fig:eHOMO_H}
\end{figure}
The KS DFT results with the standard functionals at fractional electron numbers can be easily obtained by giving fractional occupation to the HOMO.\cite{VydScuPer-JCP-07,GaiFirSta-PRL-12} As discussed in the Introduction, here we consider the challenging case of the restricted KS method, where, for singlet $N=2$ systems, as we increase the occupancy $Q$ of the HOMO orbital we should observe a jump in its energy at $Q = 1$. Notice that in the restricted KS method the conditions regarding the spin degree of freedom\cite{MorCohYan-PRL-09} are automatically fulfilled, so that the gap at $Q=1$ is the same as the ``Mott gap'' for $1/2$ spin-up and $1/2$ spin-down electrons.

The KS SCE method needs, additionally, the construction of the SCE functional for fractional electron numbers. This has been rigorously done in Ref.~\onlinecite{MirSeiGor-PRL-13}, and, in this case, corresponds to setting the co-motion function $f(r)$ of Sec.~\ref{sec:SCE} equal to
\begin{align}
	f(r)=	\begin{cases}
				N_e^{-1}[2-N_e(r)] 	& r>N_e^{-1}(2-Q)\\
				\infty			&\text{otherwise.}
			\end{cases}
			\label{eq:comffrac}
\end{align}
The physical meaning of Eq.~\eqref{eq:comffrac} is very simple: the two electronic positions are always separated by a radial distance such that the density integrates to 1 (total suppression of fluctuations),
\begin{equation}
    \int_r^{f(r)} 4\pi x^2\rho(x)\, dx=1,
\end{equation}
so that for densities integrating to less than 2 there are values of $r$ for which the second electron ``cannot enter'' in the density.\cite{MirSeiGor-PRL-13}

Figure~\ref{fig:eHOMO_H} displays $\epsilon_{\rm HOMO}$ for various approaches in the restricted KS scheme.  As observed before,\cite{MirSeiGor-PRL-13} the SCE functional shows a vertical change in the HOMO energy even in the \emph{restricted} KS approach. A sharp step, however, is only obtained with KS SCE in the extremely strong correlation (or low-density) limit,\cite{MirSeiGor-PRL-13} from which H$^-$ is still far. KS-SCE with the two local corrections exhibits the same smoothed step, but the self-interaction error leads to a non-constant HOMO energy between $0<Q<1$. The HF curve we report here has been obtained by keeping the occupancies of the two electrons equal at all $Q$. This is what it should be compared in the restricted case, and it is the situation encountered in restricted HF when stretching a bond or expanding a lattice.\cite{MorCohYan-PRL-09}
\begin{table}
	\caption{Maximum number of electrons $Q_{\rm max}$ bound in the Hydrogen nuclear potential for methods unable to bind H$^-$.}
	\centering
	\begin{tabular}{l|c|cccc}
                \hline
		& \multicolumn{1}{l|}{\emph{numerical}} & \multicolumn{4}{l}{\emph{ET-QZ3P+3diffuse}}\\
		&   KS-LDA & KS-LDA & KS-GGA & KS-metaGGA & KS-Hybrid\\
		&  (PW92) & (PW92) & (PBE) & (revTPSS) & (B3LYP)\\
		\hline
		$Q_{\rm max}$  & 1.71 & 1.71 & 1.70 & 1.73 & 1.75\footnote{by integrating over the inside region ($0<r<13$) in Fig.~\ref{fig:dens_H-_hyb}.}\\
                \hline
	\end{tabular}
	\label{tab:Nmax}
\end{table}
Finally, Fig.~\ref{fig:eHOMO_H} allows for a determination of the maximum number of electrons $Q_{\rm max}$ bound by the conventional DFT approaches. The results are compiled in Tab.~\ref{tab:Nmax}. We observe, similarly to $Z_{\rm crit}$, that the predicted value of $Q_{\rm max}$ is insensitive to the level of approximation of the standard functionals, further supporting the idea behind the model potential of Ref.~\onlinecite{GaiFirSta-PRL-12}.

\section{Conclusions and Perspectives}
\label{sec:conclusions}
We have applied functionals based on the exact strong-coupling limit of DFT to the loosely bound negative ions of the He isoelectronic series, which are a prototypical case for the delicate physics of anions and radicals.
Whereas standard DFT functionals either do not bind anions or bind them with unphysical long-range features in the charge density,
the functionals based on the strictly-correlated-electrons have a rigorous tendency to overbind that can be mitigated by local corrections.
This shows that the SCE functional and its corrections are able to capture many-body effects radically different than the ones described by the standard functionals, although improvements are still needed. In particular, one should aim at building corrections based on correlation kinetic energy effects \cite{MalMirGieWagGor-PCCP-14} and/or on exact exchange.\cite{MalMirGieWagGor-PCCP-14} 

Besides improving the accuracy of the functionals based on SCE, the challenge for the future is also to implement SCE physics into routinely applicable approximations. This can be done by either developing algorithms to evaluate the exact SCE functional exploiting its formal similarity to an optimal transport problem,\cite{ButDepGor-PRA-12,CotFriKlu-CPAM-13} as in the pilot implementation of Ref.~\onlinecite{MenLin-PRB-13}, or by constructing new approximations based on the idea of co-motion functions, {\it i.e.}, by trying to build approximate and simplified co-motion functions. These, in turn, could be used in a local interpolation along the adiabatic connection that preserves size consistency.\cite{MirSeiGor-JCTC-12}

Finally, our study also provides reference data for the anions of the He isoelectronic series close to and at the quantum phase transition that can be valuable to test the accuracy of new DFT approximations (see, e.g., Ref.~\onlinecite{BleHesUmrGor-PRA-13} which presents correlation potentials from RPA approaches that are good approximations to the true correlation potential).

\acknowledgments
We are very grateful to Erik van Lenthe for his assistance with the ADF package, and to A. J. Cohen and P. Mori-Sanchez for pointing out an error in Fig.~\ref{fig:eHOMO_H} concerning the Hartree-Fock curve. This work was supported by the Netherlands Organization for Scientific Research (NWO) through a Vidi grant and by the NSF through grant DMR-0908653.

\end{document}